\documentclass[mathleft]{an}
\usepackage{graphicx}
\usepackage{euscript}
\usepackage{times}   %%% font selection 
\begin{document}

% The following seven commands are intended for editorial usage and should be ignored by
% the author(s).
\Pagespan{001}{}% Document's page range. 
% If second parameter is left empty, the last page is computed automatically.
\Yearpublication{2011}%
\Yearsubmission{2011}%
\Month{11}%   
\Volume{999}%  
\Issue{88}% 
% \DOI{This.is/not.aDOI}% 

\title{Modified methods of stellar magnetic field measurements}
\author{A.~F.~Kholtygin\inst{1}
           \fnmsep\thanks{A.~F.Kholtygin:\email{afkholtygin@gmail.com}\newline}
}

\titlerunning{Measurements of stellar magnetic fields}
\authorrunning{A.F.Kholtygin}
\institute{
Astronomical Institute of Saint-Petersburg University, Russia
}

\received{~~ ~~~ 2012}
\accepted{~~ ~~~ 2012}
\publonline{later}

\abstract{The standard methods of the magnetic field measurement, based on an analysis of the relation between 
          the Stokes $V$-parameter and the first derivative of the total line profile intensity, 
          were modified by applying a linear integral operator $\hat{L}$ to the both sides of this relation.
          As the operator $\hat{L}$, the operator of the wavelet transform with DOG-wavelets is used. 
          The key advantage of the proposed method is an effective suppression of the noise contribution to 
          the line profile and the Stokes parameter $V$.
          The efficiency of the method has been studied using the model line profiles with various noise contributions. 
          To test the proposed method, the spectropolarimetric observations of the A0-type star $\alpha^2\,$~CVn, young O-type 
          star $\theta^1$ Ori C and A0 supergiant HD~92207 were used. 
          The longitudinal magnetic field strengths for these stars calculated by our method appeared to be in a good agreement 
          with those determined by other methods. 
         }

\keywords{stars: magnetic fields -- methods: analytical }

\maketitle

%-----------------------------------------------------------------------==1
\section{Introduction}

Measurements of stellar magnetic fields are based main\-ly on polarization observations. In 1947 Babcock 
detected for the first time the dipole stellar magnetic field of 78 Vir with the surface polar field of about 
1500 G~(\cite{Babcock1947}). 
Since that time different methods of stellar magnetic field measurements were developed. 
These methods were used for measuring the magnetic fields of about thousand stars of various spectral types 
from young T~Tau and Herbig Ae/Be stars to red giants (\cite{Bychkov-2009}). 

Methods of the magnetic field measurements are mainly based on the determination of the value of 
the displacement $\overline{\Delta\lambda}_B$ between left ($L$) and right ($R$) circular polarized 
Zeeman components of spectral lines in a presence of the magnetic field.  
Direct measurements of the value of $\overline{\Delta\lambda}_B$ are possible mostly in the case of 
spectra with sharp lines or for strong fields with magnetic induction $B$ larger than 1 kG and 
also for best quality spectra with signal-to-noise ratio $S/N\ge 1000$. Such good quality spectra can be obtained mainly 
for the brightest stars. Detection of the magnetic field of the early-type stars with wide spectral lines and moderate 
values of the magnetic field in a few hundred gauss seems to be very difficult. 

To increase the effective $S/N$ a multiline technique is usually used to extract the information 
from many spectral lines. The multiline technique is most effective for stars with a large
number of lines in their spectra. For example, for A and late type stars weak magnetic fields $B$ 
of about 1~G can be detected with 3$\,\sigma$ confidence (\cite{Petit-2010}).

   At the same time, for early-type OB and Wolf-Rayet (WR) stars with broad lines in their spectra, 
the measurement of moderate magnetic fields $B \le 100\,$G presents a very serious problem.
Moreover, the wider lines in the stellar spectra are the less effective are the measurements of a magnetic 
field of the star. This means that for different types of stars various methods of the magnetic field measurements have to be used. 

An analysis of the signal in the Stokes parameter $V$ is ordinary used to measure the magnetic fields. 
The ratio of the Stokes parameter within the spectral line, 
$V(\lambda)$, to the line profile intensity $I(\lambda)$ in the unpolarized spectrum,
can be presented as (see, e.g., \cite{Landstreet1982}, \cite{Hubrig-2006}): 
\begin{eqnarray}  %Eq.1
\label{Eq.V_dI/dlambda}    
\frac{V(\lambda)}{I(\lambda)} = - \overline{\Delta\lambda}_B \, \frac{1}{I(\lambda)} \frac{d I(\lambda)}{d\lambda} = \nonumber  & \\
-{\cal K}_0\, g_{\mathrm{eff}} \, \lambda_0^2\, \left<B_l\right> \, \frac{1}{I(\lambda)} \frac{d I(\lambda)}{d\lambda} \, .
\end{eqnarray}
where $g_{\mathrm{eff}}$ is the effective Land\'e factor for the line, 
$\left<B_l\right>$ is the mean longitudinal magnetic field, 
$\lambda_0$ is the central wavelength for the line, the 
coefficient ${\cal K}_0 = {e}/{4\pi m_{\mathrm{e}} c^2} =$  $4.6686 \cdot 10^{-13}  \:  [\mbox{\AA}^{-1}\mbox{G}^{-1}]$.
Here  $e$ is the electron charge, $m_{\mathrm{e}}$ is the mass of the electron and $c$ is the speed of light. 

Based on Eq.~(\ref{Eq.V_dI/dlambda}) two approaches, differential and integral ones are used to measure the stellar 
magnetic field. In the first approach the value of $\left<B_l\right>$ is determined as 
a coefficient of the linear regression for  Eq.~(\ref{Eq.V_dI/dlambda}). 
This regression method was described in detail by~\cite{Bagnulo-2002}. In the paper by~\cite{Bagnulo-2006} 
the estimations of the $\left<B_l\right>$ errors for the regression method are given.  

Hereinafter we will call the differential approach as differential method (DM) of the magnetic field measurement. 
The first line profile derivative $dI/d\lambda$ is calculated numerically 
(see, for example, Eq.\,(4) in the paper by Bagnulo et al. 2006), 
that leads to large errors in its value due to noise contribution in the line profile 
(e.g., \cite{Hubrig-2006}). Generally a large number of spectral lines have to be used in the DM to reach a better accuracy.

In the integral approach (integral method, IM) the first moments from the left and right sides of 
Eq.\,(\ref{Eq.V_dI/dlambda}) are calculated. For convenience, instead of the wavelength $\lambda$, 
the Doppler velocity shift from the line center $w=(\lambda-\lambda_0)/c$, is used, where $\lambda_0$ is the central 
wavelength of the line. 
Then 
\begin{eqnarray}  %Eq.2
\label{Eq.V_dI/dlambda_int}
\alpha_1(V) = \int\limits_{-\Delta w}^{\Delta w} w V(w) dw   = \nonumber  & \\ 
- {\cal R}_0 \lambda_0 g_{\mathrm{eff}} \left<B_l\right> 
\int\limits_{-\Delta w}^{\Delta w} \left[I_{\mathrm{c}}-I(w)\right] dw & \, . 
\end{eqnarray}
Here $\alpha_1(V)$ is the first moment of the Stokes parameter $V$, $I_{\mathrm{c}}$ is the intensity of the 
continuum, $I(w)$ is the continuum normalized intensity for velocity $w$. 
The value of $\Delta w$ can be taken as about of 2-3 of the line FWHM. 
The coefficient ${\cal R}_0 =1.3996 \cdot 10^{-7} \,  [\mbox{km/s}\,\mbox{\AA}^{-1}\mbox{G}^{-1}]$. Integral 
$W_{\mathrm{line}}\!\!=\!\!\!\!\int\limits_{-\Delta w}^{\Delta w}\!\left[I_{\mathrm{c}}-I(w)\right] dw\,$ is  
the equivalent width of the line in km/s. 
It should be noted that the DM is typically used for spectra with unresolved spectral lines, while the 
IM is used for spectra with resolved spectral lines. 
                                                                               
Using the moments of the line profiles of different orders in integral light and  the Sto\-kes parameter $V$ allows us 
to determine not only the mean longitudinal magnetic field $\left<B_l\right>$ but also the field crossover and  
the mean quadratic magnetic field (\cite{Mathys-1995, MathysHubrig1997}). 

\cite{Semel1989} and~\cite{SemelLi1996} used the multiline technique to reach the best accuracy of the magnetic field measurements.   
They proposed a plain averaging of the polarization measurements of many spectral 
lines to detect and analyze magnetic fields ({\it line addition}) method. 

\cite{Donati-1997} improved the line co-addition procedure and had introduced the least-squares deconvolution  (LSD) method. 
This method assumes that all spectral lines can be described by a scaled basic line profile. They also proposed that 
overlapping lines add up linearly. These assumptions led to a simplified description of the intensity
and circular polarization observations in terms of the convolution of the known line mask with an unknown line profile. 
The approximation is used to reconstruct the average line shape and the LSD profile of the Stokes parameter~$V$. 
\cite{Kochukhov-2010} extended the LSD approach to the analysis of circular and linear polarization in the 
spectral lines for the case when the shape of lines can be described by different functions. 
Alternative multiline Zeeman signature was introduced by Semel et al. (2009) and Ramirez
Velez et al. (2010). They have used the line addition technique to show how to extract
a Zeeman signature for any of the Stokes parameters. 
These techniques, apart from being applicable to any state of polarization, are model independent. 

For an effective realization of different methods of the magnetic field measurements, the polarimetric spectra with 
a rather high $S/N$ and with large number of lines in their spectra are usually required. However, application of 
the above mentioned methods to OB and WR stars displaying smaller number of very broad spectral lines is 
not simple (e.g., \cite{Kholtygin-2011b}). 

On the other hand, it is well known, that smoothing the line profiles with different filters allows one to 
suppress the noise contribution in the line profiles and to reveal the weak line profile details 
(\cite{Kholtygin-2003,Kholtygin-2006}). We suggest in this work that a certain kind of smoothing of the Stokes parameter $V$ 
enables one to achieve a larger efficiency of the various multiline methods for the magnetic field determination in upper 
main-sequence stars  and lets us to detect the magnetic fields for stars with wide and noisy line profiles. 

The modification of the standard procedures of the magnetic field measurements is described in the present paper. 
It is based on the application of the linear integral transforms to both sides of Eq.~(\ref{Eq.V_dI/dlambda}) or
Eq.~(\ref{Eq.V_dI/dlambda_int}). The transform with the DOG wavelet is used to smooth the Stokes parameter V and line profile derivative.  
The proposed approach was partly outlined by \cite{Kholtygin-2011a}.

Our paper is organized as follows. In Sect. 2 we present the main relations and basic assumptions. 
The different presentations of the linear integral transform of the Stokes parameter $V$ is considered. We examine the application of the
proposed method for the analysis of the arbitrary number of lines and spectra in Sect. 3. In this section the method is also tested for 
the different kinds of model line profiles. The application of the method to the archival observations is outlined in Sect.~4. 
In the last section we summarize the main results.

%*****************************************************************************************************************==2.
\section{Mathematical basis of the proposed methods}
\label{s.MathBasis}

%*****************************************************************************************************************==2.1.
\subsection{Main relations}
\label{s.MainRel}

\noindent
A profile of an arbitrary line in the stellar spectra as a function of a Doppler displacement $w$ 
can be presented as:
\begin{equation} %Eq.3
\label{Eq.ProfLineI}
I(w) = I_c + {\cal Z}(w) + {\cal N}(w) \, .
\end{equation}
Here  $I_c$ has the same meaning as in Eq.~(\ref{Eq.V_dI/dlambda_int}), ${\cal Z}(w)$ is a function describing the net 
line profile and ${\cal N}(w)$ is the noise contribution. Hereafter we assume that the net line profile 
${\cal Z}(w)$ is stable over the time of observations and the amplitude of the noise component is 
determined by the noise of the detector.

Similarly, the profiles of (L) and (R) polarized components of the line are determined by the formulas: 
\begin{equation}                %Eq.4
\label{Eq.ProfLineLRstar}
\left\lbrace
\begin{array}{cc}
I_{\mathrm{L}}(w)& = I^c_{\mathrm{L}} + Z_{\mathrm{L}}(w) + {\cal N}_{\mathrm{L}}(w)  \, , \\
I_{\mathrm{R}}(w)& = I^c_{\mathrm{R}} + Z_{\mathrm{R}}(w) + {\cal N}_{\mathrm{R}}(w)  \, ,  
\end{array}
\right.
\end{equation}
where symbols have the same meaning as in Eq.~(\ref{Eq.ProfLineI}), but for left-hand (L) and right-hand (R) 
polarized components separately.
The line profiles are assumed to be normalized at the continuum level. For the continuum normalized 
spectra $I_c = I^c_{\mathrm{L}}=I^c_{\mathrm{R}}= 1$. 

As a next step, we suppose that the noise components ${\cal N}_{\mathrm{L}}(w)$  and
${\cal N}_{\mathrm{R}}(w)$ are the independent random values.  
The splitting $\overline{\Delta w}_B$ of the line in the Doppler velocities space is
\begin{equation}         %Eq.5
\label{Eq.ZeemanDeltaV}
\overline{\Delta w}_B = {\cal R}_0\, g_{\mathrm{eff}} \,\lambda_0 \, \left<B_l\right> \quad [\mbox{km/s}] ,
\end{equation}

With the line intensities $I_{\mathrm{L}}(w)$ and $I_{\mathrm{R}}(w)$ for (L) and (R) polarized components 
the Stokes parameter $V$ and the total line intensity $I$ 
\begin{equation}      %Eq.6
\label{Eq.ProfLineLRstarW}
\left\lbrace
\begin{array}{cc}
       V(w) = \frac{1}{2} \left[\vphantom{\frac{1}{2}}I_{\mathrm{L}}(w)  - I_{\mathrm{R}}(w)\right] \, , \\
I(w) = \frac{1}{2} \left[\vphantom{\frac{1}{2}} I_{\mathrm{L}}(w) + I_{\mathrm{R}}(w)\right]
\, . 
\end{array}
\right.
\end{equation}
For the small value of $\overline{\Delta w}_B$, analogously to Eq.~(\ref{Eq.V_dI/dlambda}), we can write   
\begin{eqnarray}  %Eq.7
\label{Eq.V_dI/dw}
&V(w)\! = \!  - \overline{\Delta w}_B  \frac{d I(w)}{dw} \! = \! & \nonumber \\
&-{\cal R}_0\, g_{\mathrm{eff}}  \lambda_0\, \left<{B_l}\right>  \frac{d I(w)}{dw} . & 
\end{eqnarray}    
Eqs.~(\ref{Eq.ZeemanDeltaV})-(\ref{Eq.V_dI/dw}) are valid exactly only for local line profiles. 
In the simplest field configurations as the oblique rotator model there is the simple connection between the 
$\left<{B_l}\right>$ and the polar field strength \cite{Preston1967}.
In the case of more complicated field configurations the connection between the value of $\left<{B_l}\right>$ and the 
global stellar surface magnetic field distribution is not evident. 
The analysis of the longtime series of Stokes IQUV profiles and some kind of the 
Magnetic Doppler Imaging technique (e.g., Kochukhov-2004) can be used to restore this distribution.

%************************************************************************************************************************==2.2.
\subsection{Linear transform of the Stokes parameter~$V$}
\label{s.LinTransfV}
\noindent

For OB stars with typical values of $\left<B_l\right>=100-300\,$G the line shift $\overline{\Delta w}_B$ is very small,
and for noisy line profiles the derivative $dI/dw$ is calculated with large errors.  
For this reason Eq.~(\ref{Eq.V_dI/dw}) should be modified to obtain the more reasonable estimation of the
extremely small difference between (L) and (R) components. 
Applying the operator $\hat{L}$ of the arbitrary linear integral transform to the left and right sides of 
Eq.~(\ref{Eq.V_dI/dw}) we obtain: 
\begin{equation}
\label{Eq.L_BabcockV_W}
       {V}_L(w) = \hat{L}\left[V(w)\right] = \left<B_l\right> J_L(w) \, ,
\end{equation}
where ${V}_L(w)$ is the Stokes parameter $V(w)$, which is smo\-ot\-hed using the operator $\hat{L}$ and 
the integral 
\begin{equation}
\label{Eq.Jfunct_W}
 J_L(w) = -{\cal R}_0\, g_{\mathrm{eff}} \, \lambda_0\, 
\hat{L}\left[\frac{dI(w)}{dw}\right] \, ,
\end{equation}                                                          
Using Eq.~(\ref{Eq.ProfLineLRstar}) and neglecting the terms which are proportional to a small value 
$\overline{\Delta w}_B^2$ we find 
\begin{eqnarray}
\label{Eq.V_L(w)}
J_L(w)\! = \! -{\cal R}_0\, g_{\mathrm{eff}} \, \lambda_0\, \left( \hat{L}\left[\frac{dZ(w)}{dw}\right] \right. & \nonumber \\
 + \left. \frac{1}{2} \, \hat{L}\left[\vphantom{\frac{1}{2}}{\cal N}_{\mathrm{L}}(w) 
\! + \!  {\cal N}_{\mathrm{R}}(w)\right] \right) & \, .
\end{eqnarray}                                                          

The main advantage of the proposed procedure lies in a possibility to provide an effective suppression of the noise contribution 
in the line profile. 
If we use a suitable form of the linear operator $\hat{L}$ then the value of the noise component 
$\hat{L}\left[{\cal N}_{\mathrm{L}}(w) + {\cal N}_{\mathrm{R}}(w)\right]$ becomes smaller. 
It allows us to improve the accuracy of the magnetic field determination.

%-----------------------------------------------------------------------------------------------------==2.3.
\subsection{Smoothing the Stokes parameter $V$ with Gaussian filter of a variable width}

\noindent
As a possible form of the operator $\hat{L}$ we will consider a set of operators 
$\hat{L}^k_G$ of the convolution of the source function $f(x)$ with 
the family of functions ${\cal G}_k(x/S)$ having a variable width~$S$, where
\begin{equation}
\label{Eq.Gk}
{\cal G}_k(x) =  \frac{(-1)^{k}}{\sqrt{2\pi}} \frac{d^k}{dx^k} \left[
\exp\left(\frac{-x^2}{2}\right)\right] \, .
\end{equation}                                                          
with $k=0,1,2,\,\dots\,$. Coefficients in Eq.~(\ref{Eq.Gk}) are selected in such a manner that at $k=0$ the function ${\cal G}_0(x)$ 
is the unbiased Gaussian function with a width $S=1$. 
At $k\ge 1$ the function ${\cal G}_k(x)$ is proportional to DOG-wavelet $\psi_k(x)$ with an index $k$ (e.g., \cite{DeMoortel-2004}). 

With different values of the index $k$ one can construct the various approaches to measure the mean longitudinal  
magnetic field $\left<B_l\right>$. In the present paper we consider only the simplest case $k\!=\!0$. 
Applying the operator $\hat{L}^0_G$ to the left and right sides of equation~(\ref{Eq.V_dI/dw}) we obtain:
\begin{eqnarray}
\label{Eq.LG0_V_dI/dw}
&V(w,S) = \frac{1}{\sqrt{2\pi}} \times & \nonumber \\ 
& \int\limits_{-\infty}^{\infty}{\mathrm{e}}^{-\frac{1}{2}\left(\frac{w-x}{S}\right)^2} V(x) dx 
=\left<B_l\right> J(w,S) \, , \qquad &
\end{eqnarray}
where  
\begin{eqnarray}
\label{Eq.JG0funct_W}
J(w,S) =  \frac{{\cal R}_0\, g_{\mathrm{eff}} \, \lambda_0}{S^2\sqrt{2\pi}}\,  \times \nonumber & \\
 \int\limits_{-\infty}^{\infty} (w-x)\,{\mathrm{e}}^{-\frac{1}{2}\left(\frac{w-x}{S}\right)^2}
 \left[\vphantom{\frac{1}{1}}I(x)-I_c\right] dx  \, .  &
\end{eqnarray}                                                          
A value of ${V}(w,S)$ in Eq~(\ref{Eq.LG0_V_dI/dw}) is the Stokes parameter $V$, which is smoothed with the Gaussian filter 
of the variable width~$S$, while the function $J(w,S)$ in Eq~(\ref{Eq.JG0funct_W}) is proportional to the wavelet transform of 
the net line profile $[I(x)-I_c]$ with the WAVE wavelet $(2\pi)^{-1/2} x \exp(-x^2\!/2)$ and the scaling parameter $S$.
Hereafter we will use Eq.~(\ref{Eq.LG0_V_dI/dw}) to determine the value of the magnetic field $\left<B_l\right>$.

%---------------------------------------------------------------------------------------------==2.4. 
\subsection{Modified differential and integral methods of the field determination}

In the modified differential method (MDM) we use directly Eq.~(\ref{Eq.LG0_V_dI/dw}). 
The value of $\left<B_l\right>$ is derived by  the least  squares method as linear fitting of ${V}(w,S)$ vs. $J(w,S)$. 

The modified integral method (MIM) is also based on Eq.~(\ref{Eq.LG0_V_dI/dw}). 
Multiplying the left and right sides of this relation by the Doppler velocity shift $w$ and integrating 
over the line profile we obtain:
\begin{eqnarray}
\label{Eq.Vts0-Bmean_W_int}
M_{V} =
\int\limits_{-\Delta w}^{\Delta w}\!\! w\,V(w,S)  dw = & \nonumber \\ 
\left<B_l\right> \int\limits_{-\Delta w}^{\Delta w}\!\! w\, J(w,S) dw  
= \left<B_l\right> M_{J} \, .  &
\end{eqnarray}
Here ${M}_{V}$ and ${M}_{J}$ are the first moments of the Stokes parameter $V$ and the parameter $J$, respectively. 
The limits of integration $-\Delta w$ and $\Delta w$ in Eq.~(\ref{Eq.VItot_int}) are 
the same as in Eq.~(\ref{Eq.V_dI/dlambda_int}).
Eq.~(\ref{Eq.Vts0-Bmean_W_int}) is used to find the value of $\left<B_l\right>$ by the integral method.  
Results of our calculations show that the differences between the values of $\left<B_l\right>$ obtained using
both MDM and MIM in most cases are not very significant. 

%####################################################################=Fig.1
\begin{figure}[!ht]
\centering
\includegraphics[width=78mm]{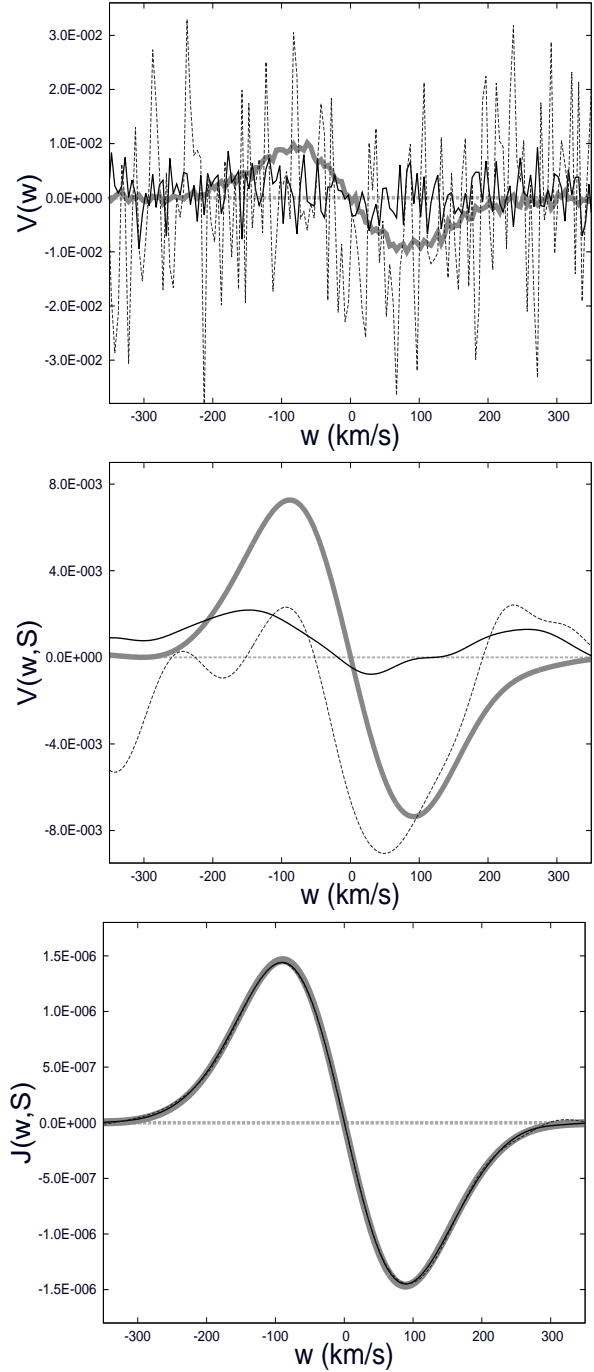}
\caption{\small {Top  panel:} the unsmoothed Stokes parameter $V(w)$ for the line HeI$\,\lambda\,4921.9\,$ with parameters  
                $Z_{0}= -0.35$, $\sigma_{w}=80\,$km/s. Thick grey solid line corresponds to $S/N=1000$,
                thin solid line comply with $S/N=200$, while dashed one corresponds to $S/N=50$.
                {Middle Panel}: the smoothed Stokes parameter $V(w,S)$, where the filter width $S=40\,$km/s  
                for $S/N=1000$ (thick grey solid line), $S/N=1000$ (thin solid line) and $S/N=50$ (dashed line).
                {Bottom panel:} the same as in the middle panel but for the parameter $J(w,S)$.
        }
\label{Fig.Fig1}
\end{figure}
%**************************************************************

%---------------------------------------------------------------------------------------------==2.5.
\subsection{Testing the method for the single line model profiles}

To test the method we use the model profiles of $L$ and $R$ com\-ponents of line calculated for a fixed input model 
field $B_{\mathrm{inp}}$. The normalized Gaussian profile for the shape of the spectral lines is used: 
\begin{equation}
\label{Eq.ProfModelNormL}
\!\!\left\lbrace
\begin{array}{ccc}
I_{\mathrm{L}}^{\mathrm{mod}}(w) & \!\!=\!\! & I_c \!\!+\!\! Z_{0} 
  \mathrm{e}^{-\frac{1}{2}\left(\frac{w - \overline{\Delta w}_B}{\sigma_w}\right)^2} \!\!+\!\! N(0, s_N)  ,  \\
I_{\mathrm{R}}^{\mathrm{mod}}(w) & \!\!=\!\! & I_c \!\!+\!\! Z_{0} 
  \mathrm{e}^{-\frac{1}{2}\left(\frac{w + \overline{\Delta w}_B}{\sigma_w}\right)^2} \!\!+\!\! N(0, s_N)  ,
\end{array}
\right.
\end{equation}
where $Z_{0}$ is the line depth at $\lambda=\lambda_0$. The value of $\overline{\Delta w}_B$ is determined by 
Eq.~(\ref{Eq.ZeemanDeltaV}), and $\sigma_w$ is the width of the model profile in km/s. $N(0, s_N)$ is the normally distributed 
random value with zero mathematical expectation and the standard deviation $s_N = (S/N)^{-1}$, where $S/N$ is 
the signal-to-noise ratio in the spectral region where the considered line is located.

The unsmoothed model profile of the Stokes parameter $V$ for the line He\,I$\,\lambda\,4921.9$ with parameters 
$Z_{0}\!=\! -0.35$, $\sigma_{w}=80\,$km/s and spectral resolving power $R=60000$ for different $S/N$ values is given 
in Fig.~\ref{Fig.Fig1} (top panel) for the illustratation. The smoothed Stokes parameter $V(w,S)$ stro\-ngly depends 
on the $S/N$ (Fig.~\ref{Fig.Fig1}, middle panel). 
In the bottom panel in Fig.~\ref{Fig.Fig1} the parameter $J(w,S)$ for the same signal-to-noise ratios is given. It is evident 
that although at $S/N=50$ the noise contribution strongly distorts the line profile and the Stokes parameter $V$, 
the value of $J(w,S)$ remains practically unchanged.  The calculations show that for all scales $S$ in the interval of 
$10\,{\mathrm{km/s}}\le S\le 160\,{\mathrm{km/s}}$ the parameter $J$ weakly depends on the $S/N$.

%xxxxxxxxxxxxxxxxxxxxxxxxxxxxxxxxxx%%%%%%%%%%%%%%%%%%%%%%%%%%%%%%%%%%%%xxxxxxxxxxxxxxxxxxxxxxxxxxxx=Fig.2
\begin{figure}[ht!]
\includegraphics[width=78mm]{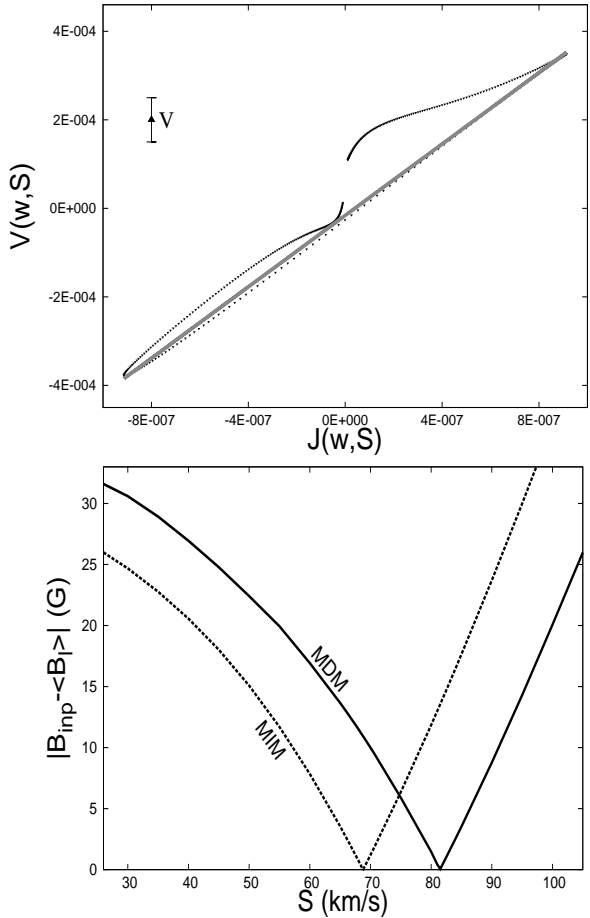}
  \caption{\small {Top panel:} dependence of Eq.~(\ref{Eq.LG0_V_dI/dw}) of the smoothed Stokes parameter $V(w,S)$ vs.  $J(w,S)$
                  for model profile of the HeI$\,\lambda\,4921\,$ at the scale $S=80\,$km/s (filled triangles) 
                  and the value of $B_{\mathrm{inp}}=400\,$G. 
                  Grey solid line shows the linear regression line for this dependence. 
                  The standard error of the parameter $V(w,S)$ is given.
                  {Bottom panel:} the absolute values of the difference
                  $|B_{\mathrm{inp}} - \left<B_l\right>|$ for the MDM (solid line) and the MIM (dashed line).
          }
\label{Fig.Fig2}
\end{figure}
%---------------------------------------------------------------

Let us to apply the MDM to the model profile of the line HeI$\,\lambda\,4921\,$  
for the input value of $B_{\mathrm{inp}}=400\,$G\ and $S/N= 800$. 
Approximating\ the dependence of the smoothed parameter $V(w,S)$ on $J(w,S)$ at the filter 
width $S=80\,$km/s and using the standard formulas of the least-squares method (e.g., \cite{Brandt1970}), we obtain 
$\left<{B_l}\right>=398\pm 18\,$G (see. Fig.~\ref{Fig.Fig2}, top panel) in a good agreement with input field value. 
Here the value of $\sigma_{\mathrm{fit}}=18\,$G is the error of the of least-squares method approximation.  
In order to reach the best fit, the contribution of the line wings is ignored when the net line intensity is close to the noise 
level. In the top panel the standard error of the smooth Stokes parameter $V(w,S)$ is plotted. The errors of the parameter 
$J(w,S)$ are too small (about of 0.01\%) to be drawn. 

Results of applying both the MDM and the MIM methods for the magnetic field determination depend on the value of 
scaling parameter $S$. Below we discuss how to choose the optimal value of $S$. 
Suppose that we vary the scaling parameter $S$ for some fixed line profile. Then, the optimal value of $S$ 
corresponds to the minimum of the absolute value of the difference $|B_{\mathrm{inp}} - \left<B_l\right>|$  between input 
field $B_{\mathrm{inp}}$ and its fitted value $\left<B_l\right>$. 
In Fig.~\ref{Fig.Fig2} (bottom panel) this difference is plotted for the MDM (solid line) 
and the MIM (dashed line). We see that the optimal value  $S_{\mathrm{opt}}\approx 81\,$km/s for the MDM,  
while $S_{\mathrm{opt}}\approx 69\,$km/s for the MIM. 
 
It means that $S_{\mathrm{opt}}\approx 0.4\, \mathrm{FWHM}$ both for the MDM and the MIM. Moreover, the minimum of the fit error 
$\sigma_{\mathrm{fit}}$ for the MDM is achieved at $S=78\,$km/s, which is close to values of $S_{\mathrm{opt}}$ both for the MDM and the MIM. 
These relations between $S_{\mathrm{opt}}$ and $\mathrm{FWHM}$ are also held for other   
parameters of the model line profiles. The following procedure can be proposed to select the optimal value of the scale $S$. 
Firstly, we vary the parameter $S$ in the interval $[0.3, 0.5]$ of FWHM.
Secondly, we are looking for the value of the scale $S$, which gives us the minimal error of the least-square fit of the 
$V(w,S)$ vs. $J(w,S)$ dependence. And finally, we choose this value of the scale $S$ as its optimal value. 
Hereinafter this algorithm of the optimal value of~$S$ selection is used.
            
It has to be noted that the value of $\sigma_{\mathrm{fit}}$ is not the error of the value of $\left<{B_l}\right>$ itself as it is often proposed. 
These errors describe the measure of the inaccuracy of the linear fit of the smoothed parameter $V(w,S)$ vs. $J(w,S)$ 
dependence only. The real error $\sigma_{\mathrm{B}}$ of $\left<B_l\right>$ can be determined by the methods described in the next section. 
For the HeI$\,\lambda\,4921\,$\AA\ line profile with parameters given above it gives 
$\sigma_{\mathrm{B}}=115\,$G. This value is much larger than~$\sigma_{\mathrm{fit}}$. 

It is worth noting that the integration in Eq.~(\ref{Eq.Vts0-Bmean_W_int}) is numerical and its 
accuracy depends on a number of points of the integration $n_{\mathrm{int}}$ in the line profile. 
This number in turn is determined by the spectral resolving power $R$ and mean profile width $V_{\mathrm{line}}$ 
in the velocity space. 
For stars with profiles broadened by rotation, the value $V_{\mathrm{line}} \approx 2 V\sin i$. 

For the numerical integrations the trapezoidal met\-hod is used. Our calculations show that the accuracy of integration 
weakly depend on the chosen numerical technique.  
 
To be certain that the magnetic field determined by the MIM is realistic, the condition $n_{\mathrm{int}} \ge 6$ has to be performed. 
For the spectral resolving power $R$ the step in the wavelengths is about $\Delta\lambda=\lambda_0/R$, where $\lambda_0$ is the 
central wavelengths of the line and the corresponding value in the velocity space is $\Delta w= c/R$. 
Assuming that $n_{\mathrm{int}}>6$ we can find the minimal width of the line profile in the velocity space, 
which can provide the necessary accuracy of the magnetic field determination, 
$\displaystyle V_{\mathrm{line}} \ge n_{\mathrm{int}}\, c/R$.

It means that the MIM is most efficient for stars with wide line profiles, e.g. fast rotating OB and WR stars.  
As for the MDM, this limitation for the line profile width is not important and this method 
can be applied to stars with both wide and narrow lines.

%----------------------------------------------------------------------------------------==2.6.
\subsection{Distribution of measured values of the magnetic field}
\label{ss.MFdistr}
\noindent

As it was pointed out above, the least-squares fit of the dependence of $V(w,S)$ vs. $J(w,S)$ can underestimate 
the error $\sigma_{\mathrm{B}}$ of the measured value of $\left<B_l\right>$. 
The real value of $\sigma_{\mathrm{B}}$ can be only obtained if we know the distribution function of the 
measured $\left<B_l\right>$ values. 
The measured value of $\left<B_l\right>$, which is determined from the analysis of the line splitting, is the random value. 
It depends on the random contribution of the noise component in the line profile for given spectropolarimetric observation. 
Each new observation, even for the fixed values of $S/N$, the fixed resolving power $R$ and other fixed parameters of 
the set of the observations, gives its own value of $\left<B_l\right>$. 
To obtain the real probability distribution for $\left<B_l\right>$ we need the extremely huge number of  
observations. Happily we can estimate the distribution function using the model line profiles. 

%xxxxxxxxxxxxxxxxxxxxxxxxxxxxxxxxxxxxxxxxxxxxxxxxxxxxxxxxxxxxxx==Fig.3
\begin{figure}[h!t]
\includegraphics[width=76mm]{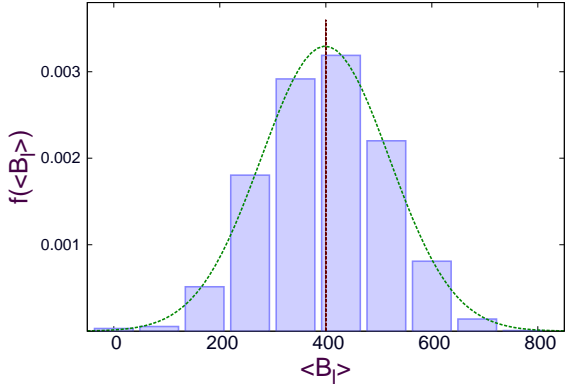}
 \caption{ \small ~The distribution function $f(\left<B_l\right>)$ of the values $\left<B_l\right>$, which is determined from the analysis 
                   of the model HeI$\,\lambda\,4921.9\,$ line profiles with parameters $R=60000$, $S/N=800$, 
                   $B_{\mathrm{inp}}=400\,$G, $Z_0=-0.35$, $\sigma_w=80\,$km/s and the scaling parameter $S=80\,$km/s. 
                   The approximation of $f(\left<B_l\right>)$ by the Gaussian is drawn by a
                   dashed line. A vertical dashed line marks the input value of $B_{\mathrm{inp}}=400\,$G.
         }            
\label{Fig.Fig3}
\end{figure}
%---------------------------------------------------------------

First, we fix the parameters  $Z_0, \sigma_w, g_{\mathrm{eff}}$ of the model profiles and 
also parameters $S/N$ and $R$, which determine the quality of the set of observations. 
Second, we simulate a large number $N_{\mathrm{sp}}\gg 1$ of the model profiles with fixed values of $S/N$ and $R$ and 
random noise contribution.
Third, we calculate the values of $\left<B_l\right>$ for all $N_{\mathrm{sp}}$ model line profiles.
And finally we divide the whole range of the calculated values of $\left<B_l\right>$ for the random noise component 
into intervals with the width of $\Delta \left<B_l\right>$. The probability distribution function $f(\left<B_l\right>)$ 
can be found from the following relation:
\begin{equation}
\label{Eq.f(B)}
f(\left<B_l\right>+\Delta\!\left<B_l\right>/2) = 
\frac{N(\left<B_l\right>, \left<B_l\right>+\Delta\!\left<B_l\right>)}{N_{sp} 
\Delta\! \left<B_l\right>}  \, ,
\end{equation}                
where $N(\left<B_l\right>, \left<B_l\right>+\Delta\!\left<B_l\right>)$ is the number of the field measurements, which gives
the value of $\left<B_l\right>$ in the interval $\left[\left<B_l\right>, \left<B_l\right>+\Delta\!\left<B_l\right>\right]$.

%---------------------------------------------------------------==Table.1
\begin{table}[h!t]
\vspace{6mm}
\centering
\caption{\small The mean longitudinal magnetic field strengths $\left<B_l\right>$ 
          and the corresponding errors $\sigma_{\mathrm{B}}$ obtained
          for different input field strengths ($B_{\mathrm{inp}}$)
          using the MDM and the MIM.  The value of the scale parameter $S=50\,$km/s.   
          }
\label{Table.ErrorsDiffInt}
\vspace{5mm}\begin{tabular}{rcccc} 
\hline
 & \multicolumn{2}{c}{\small MDM} & \multicolumn{2}{c}{\small the MIM}\\ \cline{2-5}
$B_{\mathrm{inp}},\,$G&~~~$\left<B_l\right>,\,$G~~~& $\sigma_{\mathrm{B}},\,$G & ~~~$\left<B_l\right>,\,$G~~~ & $\sigma_{\mathrm{B}},\,$G   \\
\hline
   250~~~                &  236 & 67 &   266  &  95   \\
   500~~~                &  493 & 69 &   499  &  83   \\
  1000~~~                & 1016 & 61 &  1014  &  95   \\ 
  1500~~~                & 1507 & 63 &  1504  &  97   \\ 
  2000~~~                & 2013 & 73 &  2022  &  99   \\ 
\hline
\end{tabular}
\end{table}
%---------------------------------------------------------------

In Fig.~\ref{Fig.Fig3} we present the distribution function $f(\left<B_l\right>)$ obtained by the above described method 
for the input value of $B_{\mathrm{inp}}=400\,$G from the analysis of the model profiles  of the HeI$\,\lambda\,4921.93\,$ line 
for value of $N_{\mathrm{sp}}=10^4$. 
Inspecting this figure we can conclude that the mathematical expectation of the measured field value 
$M(\left<B_l\right>)\approx B_{\mathrm{inp}}$, while the standard deviation $\sigma_{\mathrm{B}} \approx 120\,$G. 

For an illustration of the efficiency of the MDM and the MIM we present in Table~\ref{Table.ErrorsDiffInt} the average values of 
the mean longitudinal magnetic field $\left<B_l\right>$, which are determined by the MDM and the MIM  
from the analysis of the model profiles of the HeI$\,\lambda\,4921.93$ line using the numerical experiment 
described in the previous paragraphs. All model profiles are calculated for different input values of the magnetic field $B_{\mathrm{inp}}$ 
and fixed remaining  parameters of the line profiles.

Considering the results presented in Table~\ref{Table.ErrorsDiffInt} we can  conclude that the error $\sigma_{\mathrm{B}}$ of 
the determined value $\left<B_l\right>$  weakly depends on the measured value of $\left<B_l\right>$ 
itself. At the same time this value stron\-gly depends on the profile shape, $S/N$ and the spectral resolving power $R$. 
The stability of the errors in $\left<B_l\right>$ values are connected with the propagation of errors in the  noise 
contribution to the line profile (see, e.g. Bagnulo et al. 2006). 

Parameters of the model profiles are: $S/N=500$, $R=45000$, and $\sigma_w=80\,$km/s. 
From the analysis of the data presented in Table~\ref{Table.ErrorsDiffInt} we can conclude that 
the standard deviations for the the MIM are 30\% larger than for the MDM. It can be 
due to the multiplier $w$ in the expression for the first moment of the Stokes parameter $V$ in the 
Eq.~(\ref{Eq.V_dI/dlambda_int}). 
The error in the Stokes parameter $V$ grows in the line wings, where the values of $w$ are large. The contribution of 
these regions is enhanced when we multiply the value of $V$ by $w$.
The difference between the values of $\left<B_l\right>$ obtained for the MDM and the MIM, 
respectively, gives us the estimation of the accuracy of the magnetic field values.

%****************************************************************************************************************************==3. 
\section{Analysis of the multiline spectra} % 
\noindent
\label{s.MultiLineSpectra}

In this paper the multiple line spectra is treated by the same way as single lines using the {\it line addition} approximation.
Suppose that we obtain $N$ spectra for studied star and in each of them we select $m_i$ unblended spectral lines
for the magnetic field determination, where $i=1,2,\dots , N$.  The full time interval $\Delta T$ when all analyz\-ed spectra are 
obtained, must satisfy the condition $\Delta T< \gamma P_{\mathrm{rot}}$, where $P_{\mathrm{rot}}$ is the stellar rotation period and 
the parameter $\gamma \le 0.2$, 

The smoothed Stokes parameter $\EuScript{V}$ averaged over all li\-nes in the individual spectra and over all spectra  
is determined by the formula:
\begin{eqnarray}
\label{Eq.V_MultLineSp}
{\EuScript{V}} = {\EuScript{V}}(w,S) = 
\frac{\sum\limits_{i=1}^{N}\omega_i \sum\limits_{k=1}^{m_i} g_{i,k} V_{i,k}(w,S)}
     {\sum\limits_{i=1}^{N}\omega_i \sum\limits_{k=1}^{m_i} g_{i,k}} = & \nonumber \\ 
\frac{\sum\limits_{i=1}^{N}\omega_i {V}^{\mathrm{sp}}_{i}(w,S)}
     {\sum\limits_{i=1}^{N}\omega_i } \, . \qquad &
\end{eqnarray}
Here the value of $V_{i,k}(w,S)$ is the smoothed Stokes parameter $V$ for the line $k$ in spectra with the number $i$.
The value of $V_{i,k}(w,S)$ is determined by Eq.~(\ref{Eq.LG0_V_dI/dw}),  
${V}^{\mathrm{sp}}_{i}(w,S)$ is the Stokes parameter $V$ averaged over all lines in the spectra with number $i$, 
$\omega_i$ is the statistical weight of the $i$-th spectra and $g_{i,k}$ is the statistical weight of the line $k$ in the $i$-th spectra. 
The summation in Eq.~(\ref{Eq.V_MultLineSp}) is performed over all analyzed spectra and 
all selected lines.

%xxxxxxxxxxxxxxxxxxxxxxxxxxxxxxxxxxxxxxxxxxxxxxxxxxxxxxxxxxxxxxxxxxxxxxxxxxxxxxxxxxxxxxxxxxxxxxxxxxxxxxxxxxxxxxxxxxxxxxxx==Fig.4.
\begin{figure}[ht!]
\includegraphics[width=80mm]{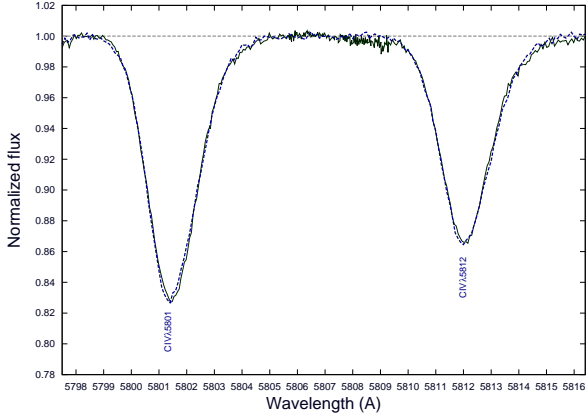}
 \caption{\small {The line profiles of the  CIV$\,\lambda\,5801,5812$\AA\ doublet in the model spectra 
                  (thick dashed line) in a comparison with those for spectra of $\lambda$ Ori A 
                  (\cite{Kholtygin-2003}, thin solid line). The parameters of the model spectra 
                  are given in the Table~\ref{Table.LamOriA_ModSpParam}. 
                 }
         }
\label{Fig.Fig4}
\end{figure}
%----------------------------------------------------------------------------------------------------------------------------

For the statistical weights of $i$-th spectrum one can use the expression $\omega_i \propto (S/N)_i^{2}$, 
where $(S/N)_i$ is the mean signal-to-noise ratio for the analyzed spectral region in the spectra with number $i$.
At the same time the value of the statistical weight of the $k$-th line in the $i$-th spectra can be evaluated via the formula  
$g_{i,k} \propto Z_0 g_{\mathrm{eff}}\left(\lambda_0^k\right)^2$, where $Z_0$, $\lambda_0$ and $g_{\mathrm{eff}}$ are the central 
depth, effective Lande factor and the central wavelength of the line $i$ in the spectra with a number $k$, respectively.
The smoothed parameter ${\EuScript{J}}(w,S)$ in the multiline approach 
can be derived in the same way as the smoothed Stokes parameter $\EuScript{V}$ in Eq.~(\ref{Eq.V_MultLineSp})
from the partial values of $J_{i,k}(w,S)$.   
Finally the mean longitudinal magnetic field $\EuScript{B}_l$ can be determi\-ned via the following relation: 
\begin{equation}
\label{Eq.VJ_multLnSp}
{\EuScript{V}}(w,S) = \EuScript{B}_l \times {\EuScript{J}}(w,S) \, ,
\end{equation}
where $\EuScript{B}_l$ stands for the mean longitudinal magnetic field in the multiline approach. 
Eq.~(\ref{Eq.VJ_multLnSp}) can be used to find $\EuScript{B}_l$ both by the MDM and by the MIM.  
In the last case we calculate the first moment from left and right sides of the Eq.~(\ref{Eq.VJ_multLnSp}). Then   
\begin{eqnarray}
\label{Eq.VItot_int}
\EuScript{M}_{V} =
\int\limits_{-\Delta w}^{\Delta w}\!\! w\,{\EuScript{V}}(w,S)  dw =  \nonumber & \\
\EuScript{B}_l \times \!\!\! \int\limits_{-\Delta w}^{\Delta w}\!\! w\, {\EuScript{J}}(w,S) dw  
= \EuScript{B}_l \EuScript{M}_{J} \, . &
\end{eqnarray}
The limits of integration in Eq.~(\ref{Eq.VItot_int}) are determined by the same manner as in Eq.~(\ref{Eq.V_dI/dlambda_int}).

The described technique for the determination of the stellar longitudinal magnetic field
can be used for an arbitrary number of the polarization spectra of different quality. 
Moreover the spectra can be located in the different spectral regions and have an arbitrary number of lines.

To illustrate the application of the modified methods of the magnetic field measurements for multiline spectra using 
Eqs.~(\ref{Eq.VJ_multLnSp})-(\ref{Eq.VItot_int}) 
we simulate the model stellar spectrum in the wavelength region $\lambda\lambda\,4000 - 6000\,$\AA.
The model spectra was created so that to be maximally close to the spectra of O8\,III star $\lambda\,$Ori~A obtained  
with the 1-m telescope of the Special Astrophysical observatory, Rusia (\cite{Kholtygin-2003}). 

For the better approach to the line profiles in the spectrum of $\lambda\,$Ori~A we use the normalized generalized 
Gaussian profile for the line shapes of the left polarized component of the lines:  
\begin{eqnarray}  %Eq.20
\label{Eq.ProfModelNormGenGauss}    
& I_{\mathrm{L}}^{\mathrm{mod}}(w) =  I_c \, + & \nonumber \\
& \left\lbrace    
\begin{array}{cc}
Z_{0}\, \mathrm{e}^{-\frac{1}{2}\left(\frac{w -\overline{\Delta w}_B}{\sigma^{\alpha}_w}\right)^{\alpha}} +  N(0, s_N), & w <0, \\
Z_{0}\, \mathrm{e}^{-\frac{1}{2}\left(\frac{w -\overline{\Delta w}_B}{\sigma^{\beta}_w}\right)^{\beta}}   +  N(0, s_N), & w \ge\ 0,  \\
\end{array} \right. &
\end{eqnarray}
% 
%

%---------------------------------------------------------------==Table.3. (new)
\begin{table}[h!t]
\tabcolsep 1mm
\vspace{6mm}
\centering
\caption{\small Parameters of lines in the model spectra of the star $\lambda\,$Ori~A in the
                interval $\lambda\lambda\,4000-6000\,$\AA
          }
\label{Table.LamOriA_ModSpParam}
\vspace{5mm}\begin{tabular}{llcrcccc} 
\hline
 Line    &~~~$\lambda_{\mathrm{lab}}$&                  &         &        &$\sigma^{\alpha}_w$&       &$\sigma^{\beta}_w$\\ 
 Name       &~~~   (\AA)             &$g_{\mathrm{eff}}$& ~~~$Z_0$&$\alpha$&  (km/s)           &$\beta$&   (km/s)         \\  \hline     
H$_\delta$  & $4101.737 $            &  $1.00$          & $-0.355$ & 1.6    &   90              &   2.15&  114             \\       
HeII4200    & $4199.87  $            &  $1.06$          & $-0.121$ & 1.7    &   60              &   1.8 &   65             \\       
H$_\gamma$  & $4340.468 $            &  $1.00$          & $-0.332$ & 1.6    &   92              &   2.1 &   85             \\       
HeI4471     & $4471.479 $            &  $1.17$          & $-0.315$ & 1.7    &   49              &   1.9 &   53             \\       
HeII4542    & $4541.591 $            &  $1.06$          & $-0.162$ & 1.6    &   58              &   2.0 &   62             \\       
HeII4686    & $4685.682 $            &  $1.07$          & $-0.201$ & 1.8    &   53              &   2.2 &   78             \\       
HeI4713     & $4713.146 $            &  $1.25$          & $-0.123$ & 2.0    &   37              &   2.0 &   43             \\       
H$_\beta$   & $4861.332 $            &  $1.00$          & $-0.352$ & 1.8    &  116              &   2.2 &   94             \\       
HeI4921     & $4921.931 $            &  $1.00$          & $-0.164$ & 1.7    &   47              &   1.8 &   43             \\       
HeI5016     & $5015.678 $            &  $1.00$          & $-0.140$ & 1.8    &   44              &   2.0 &   43             \\       
HeII5411    & $5411.521 $            &  $1.06$          & $-0.207$ & 1.7    &   58              &   2.0 &   66             \\       
OIII5592    & $5592.37  $            &  $1.00$          & $-0.155$ & 2.0    &   44              &   1.9 &   40             \\       
CIV5801     & $5801.33  $            &  $1.33$          & $-0.173$ & 1.8    &   37              &   1.9 &   52             \\       
CIV5812     & $5811.98  $            &  $1.17$          & $-0.136$ & 1.8    &   40              &   1.8 &   50             \\       
HeI5876     & $5875.966 $            &  $1.17$          & $-0.375$ & 2.6    &   70              &   1.5 &   35             \\       
\hline                                                                                                                    
\end{tabular}                                         
\end{table}
%---------------------------------------------------------------

The right polarized component can be also written using the Eq.~(\ref{Eq.ProfModelNormGenGauss}), but replacing $-\overline{\Delta w}_B$ by 
$+\overline{\Delta w}_B$. 
The parameters $\alpha$, $\sigma^{\alpha}_w$ and $\beta$, $\sigma^{\beta}_w$ describing the line shape for the violet and red 
parts of the line profile can differ. All the other parameters are the same as used in the Eq.~(\ref{Eq.ProfModelNormL}).

We fit the line profiles of 15 isolated lines in spectra of $\lambda\,$Ori~A in the range $\lambda\lambda\,4000-6000\,$\AA.
% The  fitting of of 15 isolated lines profiles is  simulated for for lambda OriA in the range lambda 4000-6000 AA. 
The list of the fitted lines is presented in the 1st column of Table~\ref{Table.LamOriA_ModSpParam}. 
The laboratory wavelengths $\lambda_{\mathrm{lab}}$ listed in the 2nd column, were taken from the catalogue by~\cite{ReaderCorliss1980}.
The effective Land\'e factors were computed thro\-ugh the classical formulas for LS-coupling (e.g.,~\cite{MathysStenflo1986}). 
The fitting parameters $Z_0$, $\alpha$, $\sigma^{\alpha}_w$, $\beta$ and $\sigma^{\beta}_w$ of all 15 lines are given in 4th\,-\,8th 
columns respectively. The quality of the fit is quite good as it shown in Fig.~\ref{Fig.Fig4}.

Three model stellar spectra with the input magnetic field value $B_{\mathrm{inp}}=500\,$G were generated for the values of 
$S/N$ = 500, 750 and 1000. The line parameters were taken from Table~\ref{Table.LamOriA_ModSpParam}. After applying 
Eqs.~(\ref{Eq.VJ_multLnSp})-(\ref{Eq.VItot_int}) the mean value of $B_l=512\pm 14\,$G for MDM and $B_l=498\pm 15\,$G for MIM 
were obtained. These values differ from the input value by less than one standard deviation. 

%*************************************************************=============================3.1. 
\subsection{Magnetic field measurements for spectra with low signal-to-noise ratio}
\label{ss.lowS/N}
\noindent

As it was mentioned above, the proposed methods seem to be the most effective for spectra of hot stars with broad lines 
and low S/N. Consider the model spectra described in the previous section in the 
region $\lambda\lambda\,4000 - 6000\,$\AA\  with lines, which are given in Table~\ref{Table.LamOriA_ModSpParam}. 
Further we assume the line profile width $\sigma_w=250\,$km/s and the input value of $B_{\mathrm{inp}}=1\,$kG, which is typical for 
star with the dipole field and the polar field value $B_p=4-5\,$kG. We also assume the relatively small spectral 
resolving power $R=15000$.

%xxxxxxxxxxxxxxxxxxxxxxxxxxxxxxxxxxxxxxxxxxxxxxxxxxxxxxxxxxxxxxxxxxxxxxxxxxxxxxxxxxxxxxxxxxxxxxxxxxxxxxxxxxxxxxxxxxxxxxxxxxxx==Fig.5
\begin{figure}[ht!]
\centering
\includegraphics[width=64mm]{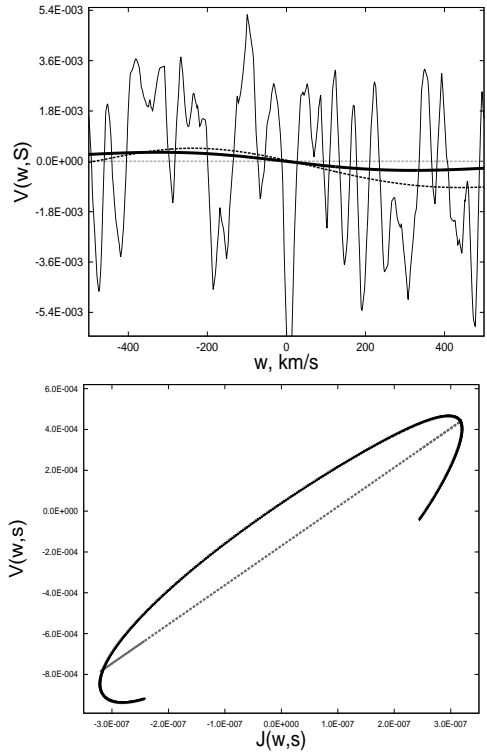}
 \caption{\small {Top panel}: the unsmoothed Stokes parameters $V(w)$ for model spectra with lines 
                 given in Table~\ref{Table.LamOriA_ModSpParam} for $\sigma_w=250\,$km/s, $S/N=300$, $R=15000$ 
                 (thin solid line) and the smoothed Stokes parameters $V(w,S)$ for the scale parameter 
                 $S=200\,$km/s (dashed line) in a comparison with the Stokes parameters $V(w,S)$ for the same spectra, 
                 but with the zero noise contribution (thick solid line).    
                 {Bottom panel}: the $\EuScript{V}(w,S)$ vs. $\EuScript{J}(w,S)$ dependence (filled triangles) and its 
                 linear fit (dashed line).
         }
\label{Fig.Fig5}
%\label{Mod.Sp.lowS/N}
\end{figure}
%----------------------------------------------------------------------------------------------------------------------------

Our calculations show that even for low $S/N\approx 300$, the longitudinal magnetic field of about of $\sim 1\,$kG 
can be detected. The procedure of the field detection for the value of $S/N=300$ is illustrated in Fig.~\ref{Fig.Fig5}. 
The determined values of $\left<B_l\right>=1203\pm 190\,$G for the MDM and $1099\pm 105\,$G for the MIM are 
in agreement at 2$\sigma$ level with the input value of $B_{\mathrm{inp}}=1000\,$G. Spectra of such quality can be 
obtained not only for galactic OB and WR stars but also for the stars of similar types in the Magellanic clouds and 
galaxies M31 and M33. It means that it is possibly to inspect the nearest galaxies for the presence of the 
magnetic early-type stars.

%*******************************************************************************************************************==3.2.
\subsection{A comparison of integral and modified integral methods} 
\label{ss.compIM-MIM}
\noindent

To test the efficiency of the proposed modified versions of integral method of the magnetic field measurement we 
consider the model spectrum in the 
region $\lambda\lambda\,4000\! - \! 6000\,$\AA\, using lines which are given in Table~\ref{Table.LamOriA_ModSpParam} and with the line widths which are 
the same as in the spectra of $\lambda\,$Ori~A, as it described in the section~\ref{s.MultiLineSpectra}. 
We fix the $S/N=400$ and the spectral resolving power $R=60000$ and compare the values of the measured magnetic field 
for different input values of $B_{\mathrm{inp}}$ obtained with the standard IM using 
Eq.~(\ref{Eq.V_dI/dlambda_int}) and by the MIM.

%-----------------------------------------------------------------------------------------==Table.3.
\begin{table}[h!t]
\vspace{6mm}
\centering
\caption{\small A comparison of the input model magnetic field values $B_{\mathrm{inp}}$ 
                with averaged over all lines values $\overline{\left<B_l\right>}$ 
                obtained by IM and MIM together with the corresponding $\sigma_{\mathrm{B}}$ 
                value of the scale parameter $S=40\,$km/s   
       }
\label{Table.MF_compIM-MIM}
\vspace{5mm}
\tabcolsep 2mm
\begin{tabular}{rcccccc} 
\hline
          &  \multicolumn{3}{c}{\small IM}    &    \multicolumn{3}{c}{\small MIM$^a$}                        \\ 
\hline
$B_{\mathrm{inp}},\,$G & $\overline{\left<B_l\right>},\,$G & $\overline{\sigma}_{\mathrm{B}},\,$G & $\sigma_{\mathrm{B}}$  & 
$\overline{\left<B_l\right>},\,$G & $\overline{\sigma}_{\mathrm{B}},\,$G & $\sigma_{\mathrm{B}}$              \\
\hline                         
%            B_IM      sig_IM      sig_IM(DF)                       B_MIM       sig_MIM    sig_MIM(DF)
 500      & $ 891$    & $480$    &   $ 550$                      & $ 706$      &  $244$   & $125$  \\ 
1000      & $ 803$    & $588$    &   $ 666$                      & $ 913$      &  $193$   & $176$  \\ 
2000      & $2718$    & $972$    &   $ 550$                      & $2223$      &  $226$   & $144$  \\ 
4000      & $4448$    & $727$    &   $ 517$                      & $4195$      &  $270$   & $161$  \\ 
8000      & $7291$    & $879$    &   $ 589$                      & $8018$      &  $183$   & $181$  \\                          
\hline                                            
% \multicolumn{7}{l}{\small $^a$calculated for $S=40\,$km/s}\\                  
\end{tabular}                                           
\end{table}
%---------------------------------------------------------------
%

The comparison of the results obtained with the IM and the MIM is presented in Table~\ref{Table.MF_compIM-MIM}. In the first 
column of the table we give the input model magnetic field $B_\mathrm{inp}$, in the 2nd and 3rd columns 
we present the magnetic fields $\overline{\left<B_l\right>}$, which are determined by the IM from the analysis of the model 
spectra and the corresponding $rms$ errors $\overline{\sigma}_{\mathrm{B}}$. 
We use the standard formulas~(\cite{Brandt1970}): 
\begin{eqnarray}
\label{Eq.BmeanSigmaB}
& \overline{\left<B_l\right>}     = \frac{1}{n}\!\sum\limits_{i=1}^{n} \left<B_l\right>_i , \nonumber & \\
& \overline{\sigma}^2_{\mathrm{B}}= \frac{1}{n-1}\!\sum\limits_{i=1}^{n} \left(\left<B_l\right>_i - \overline{\left<B_l\right>}\right)^2 . &
\end{eqnarray}
Here $\overline{\left<B_l\right>}_i$ is the mean longitudinal magnetic field derived from the analysis of the line number $i$,  
where $n=15$ lines were used to determine the magnetic field. 
In the 4th column the real error $\sigma_{\mathrm{B}}$ derived from the 
distribution fun\-ction for ${\left<B_l\right>}$ (see subsection~\ref{ss.MFdistr}). 
Last three columns give the same values as those, given in the 2nd, 3rd and 4th columns, but for the MIM. 
It is clear that the MIM method gives more exact values of measured fields (at least for the model spectra)
and smaller errors of the field determination.   

The errors $\overline{\sigma}_{\mathrm{B}}$, which are given in Table~\ref{Table.MF_compIM-MIM} are larger than 
the real error $\sigma_{\mathrm{B}}$ derived from the distribution function for ${\left<B_l\right>}$ 
(see subsection~\ref{ss.MFdistr}). 

But even if we use the error  $\overline{\sigma}_{\mathrm{B}}$ instead of $\sigma_{\mathrm{B}}$,  
the moderate fields can be easily detected by the MIM at the $3 \overline{\sigma}_{\mathrm{B}}$ level 
for $B_{\mathrm{inp}}>500\,$G. 
The large scattering of the values of $\overline{\sigma}_{\mathrm{B}}$ in the Table~\ref{Table.MF_compIM-MIM} is 
connected with the relatively low $S/N$ value.

It should be concluded that the proposed MIM of the magnetic field determination 
works better for low S/N polarized spectra comparing to the more simple IM. 
Our calculations show that the same conclusion is also valid for DM and MDM methods.

%*************************************************************==============================================4.
\section{Application to Archival Observations}
\label{s.Appl-to-obs}
\noindent               

To test the efficiency of the proposed method for determination of the stellar magnetic field we 
apply it to study the magnetic field strength of two well known magnetic stars 
$\alpha^2\,$~CVn, \,$\theta^ 1$Ori~C and AO star HD\,92207 with recently measured magnetic fields.
% star compare the measured values with those obtained by other methods.

%*************************************************************==4.1.
\subsection{$\alpha^2\,$~CVn}
\label{ss.alpha2CnVn}
\noindent

The chemically peculiar Ap star $\alpha^2\,$~CVn (HD\,112413) of the spectral class A0 is 
often used as the magnetic field standard. The effective temperature of the star is equal to $11600\pm 500\,$K 
(Kochukhov \& Wade 2010), its luminosity $\approx 10^2\,L_{\odot}$ and the rotation velocity 
$V\sin i=18\,$km/s~(\cite{Kochukhov-2002}). To test our methods of the magnetic field determination 
we used the polarization spectra of $\alpha^2\,$~CVn obtained on February 2, 2009
at the 6-meter telescope of the Russian Special Astrophysical observatory by \cite{Chountonov-2011}.  
\cite{KochukhovWade2010} obtained the parameters of the stellar magnetic field of $\alpha^2\,$ CVn 
in the context of the oblique rotator model~\cite{Preston1967} from the measurements of the magnetic field on the Balmer lines. 

%---------------------------------------------------------------==Table.4. (3)
\begin{table}[h!t]
\tabcolsep 1mm
\vspace{6mm}
\centering
\caption{\small Results of the magnetic field measurements (in G) for $\alpha^2\,$~CVn 
                for hydrogen lines by DM using Eq.~(\ref{Eq.V_dI/dlambda}), by MDM and MIM for $S=40\,$km/s 
                and obtained from the phase given by~\cite{KochukhovWade2010} 
          }
\label{Table.MF_alp2CVn}
\vspace{5mm}\begin{tabular}{ccccc} 
\hline
             &                 &               &                & Phase        \\
 Line        & DM$^a$          &MDM            &MIM             & curve$^c$   \\
\hline                                              
H$_{\gamma}$ & $-1419\pm 108$  & $-950\pm 59$  & $-1161\pm 51$  &  -          \\
H$_{\beta}$  & $ -662\pm  81$  & $-624\pm 25$  & $ -706\pm 64$  &  -          \\
 H-lines     & $ -971\pm  68$  & $-787\pm 32$  & $ -933\pm 41$  & $-890$      \\ 
\hline                                                  
\end{tabular}                                           
\end{table}
%---------------------------------------------------------------

The rotation phase $\phi$ at the time $T$ can be calculated via the formula $\phi= (T-T_0)/P_{\mathrm{rot}}$,
where the rotation period $P_{\mathrm{rot}} = 5^{\mathrm d}.46939$ (\cite{Farnsworth1932}). 
Using this relation and the Julian dates of the observations we
compute the rotation phase $\phi = 0.138$ at the mean time of observations of $\alpha^2\,$~CVn. 
For this phase the value of the longitudinal magnetic field calculated from the magnetic field phase curve 
by~\cite{KochukhovWade2010} is $\left<B_l\right> = -890\,$G. 

In Table~\ref{Table.MF_alp2CVn} we compare the listed in the 2nd column the mean longitudinal magnetic field $\left<B_l\right>$ values 
obtained via the standard relation Eq.~(\ref{Eq.V_dI/dlambda}) of the differential method, with those  
determined by the MDM and the MIM using Eqs.~(\ref{Eq.LG0_V_dI/dw}) and~(\ref{Eq.Vts0-Bmean_W_int}) for hydrogen lines (3rd and 
4th columns). The $\left<B_l\right>$ values, which are determined in the different approaches are in good agreement. The detected 
longitudinal magnetic field strengths are also close to the value $-890\,$G, obtained from 
the phase curve by~\cite{KochukhovWade2010}.

%*************************************************************====================================4.2.
\subsection{$\theta^1$ Ori C and HD\,92207}
\label{ss.tet01OriC}
\noindent

We also apply our modified methods of the magnetic field measurement also to the star $\theta^1$ Ori C, which was the first
O-type star with the detected magnetic field varying with the rotation period of 15.4 days~(\cite{Donati-2002}). 
To determine the mean longitudinal magnetic field $\left<B_l\right>$ we use the Stokes $I$ and 
$V$ profiles obtained by~\cite{Hubrig-2008} for twelve observations of $\theta^1$ Ori C distributed over the rotational period.
The observations were carried out in 2006 in service mode at the European Southern Observatory with FORS~1 mounted on
the 8-m Kueyen telescope of the VLT with GRISM 600R in the wavelength range $5240-7380\,$\AA.  
The narrowest slit width of $0''.4$ was used to obtain a spectral resolving power of $R \sim 3000$ with GRISM 600R.
The measurements were reduced in the same manner as by~\cite{Hubrig-2008} and after that were used to determine the
magnetic field by the MDM and the MIM. 

We compare the mean $\left<B_l\right>$ values for $\theta^1$ Ori~C obtained by us with those found by other authors. 
and conclude that our measurements are in a good agreement with the results by \cite{Hubrig-2008}, 
\cite{Wade-2006} and \cite{Petit-2008}.     
The largest discrepancy between the measurements obtained by~\cite{Hubrig-2008} and by us occurs only 
for MJD=54182.048 at the rotation phase 0.8777.  Our calculations give the values $455\pm 230\,$G and $421\pm 198\,$G 
for the MDM and the MIM respectively, while~\cite{Hubrig-2008} report the value $84\pm 54\,$G. 
But even in this case the data presented in the present paper is consistent with results published by \cite{Hubrig-2008} within 
$2\sigma$ interval.

According to \cite{PrzybillaNieva2011}, the abundan\-ces ratio N/C$>0.8$ in B-type mainsequence stars indicate their possible magnetic nature.
The recent determination of the atmospheric elemental abundances for the visually brightest early A0 supergiant HD\,92207 
(Przybilla et al. 2006) indicates a modest enrichment of nitrogen with the N/C abundance ratio of 0.83. It means that this star is 
a good target for searching of the magnetic field in early A-type supergiants.
The magnetic field measurements for this star were based on spectropolarimetric observations fulfiled during 2011
and 2012 with the multi-mode instrument FORS~2 installed at the 8-m An\-tu telescope of the VLT and were reported by \cite{Hubrig-2012}.
The authors reported on the detection of the longitudinal magnetic field of HD\,92207 at a significance level of more than 3$\sigma$ 
for dates $MJD=55\,688.168$ and $MJD=55\,936.341$. 

We use the observations of HD 92207 for this dates to test our modified methods of the magnetic field measurements. 
The determined by us values of $\left<B_l\right>$ are given in Table~\ref{Table.MF_HD92207}.
In the first column of the table the dates of observations is pre\-sented. In the 2nd column the line sets which are used for the field 
determinations are pointed. The set [all] includes all lines which can be used for the field determinations, while the set [hyd]  
contains only hydrogen lines. Results of the field determinations by \cite{Hubrig-2012} are presented in the 3rd column.
In 4th and 5th columns the field values obtained by MDM and MIM are given. Analysing the Table~\ref{Table.MF_HD92207} 
one can conclude that both MDM and MIM fields values are in a good agreement with those given by \cite{Hubrig-2012}.

%---------------------------------------------------------------==Table.5. (new)
\begin{table}[h!t]
\tabcolsep 1mm
\vspace{6mm}
\centering
\caption{\small Results of the mean efeective magnetic field $\left<B_{l}\right>$ measurements (in G) for HD$\,$92207 using 
                FORS$\,$2 observations by DM (Hubrig et al. 2012) and by MDM and MIM  for $S=30\,$km/s 
          }
\label{Table.MF_HD92207}
\vspace{5mm}\begin{tabular}{lrrrr} 
%----------------------------------------------------------------------------------------------------
\hline
             & Line &                &            &             \\
   MJD       & sets & DM~~           &   MDM~~    & MIM~~       \\  \hline                                              
%---------------------------------------------------------------------
 55\,688.168 &[all] & $-384\pm 42$   &$-473\pm 83$&$-477\pm 100$\\
             &[hyd] & $-402\pm 52$   &$-530\pm 68$&$-527\pm  89$\\  
%----------------------------------------------------------------------
 55\,936.341 &[all] & $ 145\pm 38$   &$ 247\pm 46$&$ 250\pm 77$ \\ 
             &[hyd] & $ 157\pm 51$   &$ 246\pm 45$&$ 247\pm 58$ \\  \hline                                                  
%--------------------------------------------------------------------------
\end{tabular}                                          
\end{table}
%--------------------------------------------------------------

Resuming the results of this section we can conclude that proposed in the present paper MDM and MIM met\-hods give a good results for 
simple fields. They also let to reach the acceptable results for the complex structure of the magnetic fields 
which are close to those obtained by other methods.

%*************************************************************==5.
     \section{Conclusion}
\noindent
\label{zakl}

In the present paper the modified methods of the stellar magnetic fields measurement are described.  
This method is ba\-sed on the application of the wavelet transform with DOG-wavelets to the integral line profiles and the 
smoothing with the Gaussian function to the Stokes parameter $V$. 
The proposed me\-thod can be used both for isolated lines in the stellar spectra and for the wide spectral regions including 
arbitrary numbers of the unblended lines, and also for a number of stellar spectra with close rotation phases.

The application of the proposed method to the model stellar spectra shows that this method is most efficient for stellar 
spectra having unblended lines with line widths lar\-ger than $30-40\,$km/s.
The main advantage of the method is the effective suppression of the noise contribution both to the line
profile in integral light and in the Stokes parameter $V$. Mo\-reover, this method can be tuned for 
the individual spectra by the selection of the optimal value of the scale $S$. 
The proposed in the present paper MDM and MIM methods are  valid for simple, large-scale fields and using them for the magnetic fields of 
complex structure may be less successful. We plan to investigate the efficiency both MDM and MIM methods for 
complex fields in following papers. 

In the paper we consider only the case $k\!=\!0$ of the smoothing the Stokes parameter $V$  with the family of functions 
${\cal G}_k(x/S)$, which are determined by Eq.~(\ref{Eq.Gk}). Accepting the value of $k\!=\!1$ we obtain the gaussian-like form 
of the smoothed function $V_1(w,S)$. It means that the case $k\!=\!1$ may be more convenient for the analysis of the polarized 
spectra with the overlapping lines. The cases $k\!>\!1$ probably are not very suitable for such analysis due to the complex 
structure of the DOG-wavelets with indexes $k\ge 2$. The proposed method for the case $k\!=\!1$ will be developed in the 
future publications.

\acknowledgements
   I am grateful to the referee for his valuable comments.
   I thank to G.~A. Chountonov  for the presentation the polarized spectra of $\alpha^2\,$~CVn 
   and N.~Rusomarov for processing this spectra with the MIDAS package. 
   I am also grateful to S.\,Hubrig for giving a possibility to use her spectropolarimetric observations of 
   $\theta^1$ Ori C and HD\,92207 to test the methods. This work was supported by a Saint-Petersburg 
   University project 6.38.73.2011.

%%%%%%%%%%%%%%%%%%%%%%%%%%%%%%%%%%%%%%%%%%%%%%%%%%%%%%%%%%%%%%%%%%%%%%%%%%%%%%%%%%%%%%%%%%%%%%%%%%%%

\end{document}